\documentclass[prl,preprint,showpacs]{revtex4}

\usepackage{amsmath}
\usepackage{epsfig}
\usepackage{rotating}

\begin{document}

\title{Metallic and semi-metallic $\langle 100 \rangle$ silicon nanowires}

\author{R. Rurali}
\author{N. Lorente}
\affiliation{Laboratoire Collisions, Agr\'{e}gats, R\'{e}activit\'{e},
             IRSAMC, Universit\'{e} Paul Sabatier,
             118 route de Narbonne, 31062 Toulouse c\'{e}dex,
             France}

\date{\today}

\begin{abstract}
Silicon nanowires grown along the $\langle 100 \rangle$-direction with
a bulk Si core are studied with density functional calculations. 
Two surface reconstructions prevail after exploration of a large fraction 
of the phase space of nanowire reconstructions. Despite their energetical
equivalence, one of the reconstructions is found to be strongly metallic
while the other one is semi-metallic. This electronic-structure behavior is
dictated by the particular surface states of each reconstruction. These results 
imply that doping is not required in order to obtain good conducting Si nanowires.
\end{abstract}


\maketitle

One-dimensional semiconductor nanostructures are attracting great 
interest for their potentially high impact in future 
molecular-electronics applications, such as nanoswitches and 
nanocontacts~\cite{appell:morales:lieber}. 
Silicon nanowires~(SiNWs) appear to be an especially appealing choice, 
due to their ideal interface compatibility with conventional Si-based 
technology~\cite{huang,lauhon}. Moreover, measuring changes of their 
conductance allows to use SiNWs as real-time label-free sensors in 
different chemical environments. The detection of NH$_3$~\cite{zhou} 
and of biological macromolecules~\cite{cui} has been reported,  
envisaging the possibility to reach the single-molecule detection 
limit~\cite{hahm}.

Recently, SiNWs of diameters below 10 nm have been synthesized.
Ma {\em et al.}~\cite{ma} have reported extremely thin SiNWs 
grown along the $\langle 110 \rangle$ lattice direction, whose diameters 
range from 1.3 to 7~nm. Previously, Holmes {\em et al.}~\cite{holmes}
obtained $\langle 100 \rangle$ and $\langle 110 \rangle$ SiNWs of
4 to 5~nm and discussed the influence of lattice orientation. Wu 
{\em et al.}~\cite{wu} have grown $\langle 110 \rangle$, $\langle 111 \rangle$
and $\langle 112 \rangle$ SiNWs down to 3~nm. Colemann 
{\em et al.}~\cite{colemann} have reported SiNWs of 3 to 5~nm diameter. 

Thus, a detailed understanding of thin SiNWs structure and of their
mechanical and electrical properties as a result of the different 
growth conditions is required. While it has been extensively demonstrated
that in H-terminated SiNWs quantum confinement induces a gap-broadening
effect~\cite{delley,read,uzi}, little is known about surface 
reconstruction of non-passivated wires and about their electronic structure.
Theoretical studies of the structure of the thinnest possible 
SiNW~\cite{menon,li} have been previously published. However, recent 
experiments have provided convincing evidence that some SiNWs 
grow around a mono-crystalline bulk Si core~\cite{zhang,ma}. Systems 
fulfilling such a requirement have been studied by Ismail-Beigi 
and Arias~\cite{arias}  ($\langle 100 \rangle$ SiNWs) and by Zhao and 
Yakobson~\cite{yakobson} ($\langle 100 \rangle$ and $\langle 110 \rangle$ 
wires). Both studies show the paramount importance of facet edges 
in wires with diameters in the nanometer range.

Thin nanowires pose fundamental problems with respect 
to their conduction properties. Recently, it has been discussed that 
doping may not be the advisable technique for tailoring nanodevice 
electrical properties, because of the expected statistical deviation of 
impurity concentration from one system to another~\cite{pikus,uzi}. 
Indeed, typical concentrations for attaining the measured SiNWs 
conductance~\cite{heath,liebercond} may well mean that no donor/acceptor 
is actually present in nm-wide SiNWs. Some of these measurements show 
that conductance is larger than expected for doped SiNW~\cite{heath} 
and the authors suggest that surface states may be responsible for the 
measured conduction.

In this Letter, we present a theoretical study of realistic 
$\langle 100 \rangle$ SiNWs, with a bulk Si core and a diameter of 
$\sim$~1.5~nm~\cite{arias,yakobson}, thoroughly exploring the phase 
space of SiNW reconstructions. We study different periodic cells,
ranging from 57 to 912 atoms, so that periodicity does not prevent us
from finding the lowest-energy reconstruction. 
The mechanical properties of the SiNWs are 
studied by computing their Young modulus and Poisson ratio.  Particularly, 
we focus our attention on the electronic properties of the wire as 
determined by the different possible lateral facet reconstructions.
We give the first description of nanowire surface states as 
they evolve from the particular surface reconstruction of the 
studied SiNW. Depending on the reconstruction, the surface state
can cross the Fermi level affecting the electrical characteristics 
of the nanowire. Hence, there is an intrinsic relation between the 
reconstruction of nanowire facets and its transport properties. This is a 
remarkable finding that shows that doping is not needed for
obtaining conducting SiNWs.

We have performed density-functional theory~(DFT) calculations with both 
a numerical atomic orbital~\cite{siesta} and a plane-wave~\cite{dacapo} 
basis set.  We have used a double-$\zeta$ polarized basis set~\cite{siesta} 
and a plane-wave energy cutoff of 20~Ry~\cite{dacapo}, with the Generalized 
Gradient Approximation~\cite{gga} for the exchange-correlation functional. 
We have studied wires in supercell geometry with 
a diameter of $\sim 1.5$~nm.  The axis periodicity will restrain the 
number of possible reconstructions and for this reason we have considered 
different supercell sizes, analyzing SiNWs of 57, 114 and 171 atoms.
The reciprocal space has been sampled with a converged grid 
of 12, 6 and 4 k-points respectively. The atomic positions have been relaxed 
until the maximum force was smaller than 0.04~eV/\AA.

The faceting geometry adopted by the wire is given by thermodynamical 
considerations~\cite{arias,yakobson,forthcoming}. On the one hand, the formation 
of \{100\} facets is favored over \{110\} facets, due the lower 
corresponding surface energy; on the other hand, facets with an even
number of atoms can dimerize, lowering their energy, and are thus
favored over facets with an odd number of atoms~\cite{ZhangCol}. 

\begin{figure}[h]
\begin{center}
\epsfxsize=7cm
\epsffile{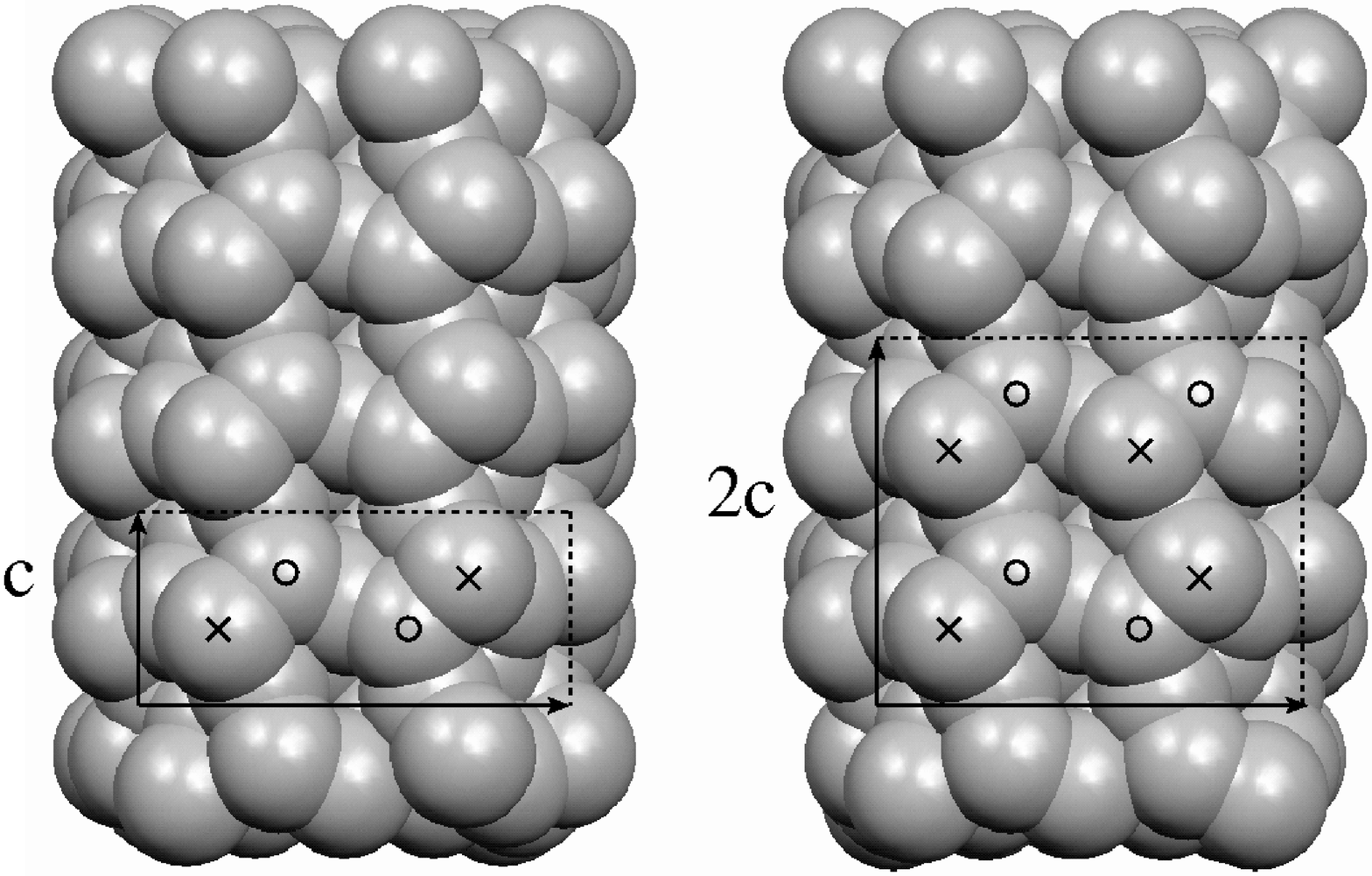}
\end{center}
\caption{ \{100\}-facet reconstructions. (a) $1c$ and (b)
           $2c$ dimerization of the surface dangling bonds.
           Lattice vectors of the facet unit cell are sketched to
           underline the different periodicity along
           $\langle 100 \rangle$ axis ($c$ is the bulk lattice
           parameter).}
\label{fig:facets}
\end{figure}

We have obtained two competing geometries for the \{100\} facets: a 
$1c$ reconstruction [see Fig.~(\ref{fig:facets}a)] and a 
$2c$ reconstruction [see Fig.~(\ref{fig:facets}b)], being $c$ the 
lattice vector along the axis of the unreconstructed wire.
The $1c$ reconstruction has the same periodicity of bulk Si, presents
a {\em trough} in the middle of the facet  
and turns out to be the most stable, with a cohesive 
energy of 3.990~eV/atom~\cite{cohesion}. This value is only 3~meV/atom 
lower than for the $2c$ reconstruction. The small energy difference 
-~probably within the accuracy of our calculations~\cite{nota,ramstad,healey}~- 
indicates that both reconstructions are likely to coexist.
The trough in the $1c$ reconstruction is determined by the buckled dimer 
sequence, which presents a double chain of {\em low} atoms [labeled 
with a circle in Fig.~(\ref{fig:facets})]; this rule is respected by 
only one of the two sides in the $2c$ reconstruction, while on the other 
side one every two dimers is {\em flipped}. In contrast to what has 
been proposed for the Si(100) surface~\cite{artacho}, none of the two 
cases shows spin polarization. 

The two arrangements do not correspond to any infinite surface 
reconstruction, even though they loosely resemble the Si(100) $c(4\times2)$ 
and $p(4\times2)$ (but not the $p(2\times2)$, for the Si(100) surface 
reconstructions see for example Ref.~\cite{ramstad}). The difference 
between surface and SiNW reconstructions stems from the lower coordination 
of the facet atoms. Between two adjacent dimers on the facet there is 
one single atom in the underneath layer, while between the corresponding 
dimers on the (100) surface there are two. This reduced coordination 
leads to a lateral shift of the dimers, increasing their packaging.

In order to extensively explore other possible reconstructions, we have 
also performed non-orthogonal tight-binding~(TB)~\cite{seifert,trocadero} 
calculations, considering supercells with a lattice parameter 
along the wire axis up to $16c$ (912 atoms). The TB structural relaxations 
are in very good agreement with the DFT results and no new reconstruction
was found. We have also taken into account other faceting arrangements 
prior to relaxations, but we confirm that the minimum energy 
configurations are obtained when \{100\} facets prevail~\cite{arias,yakobson,forthcoming}. 
Second-neighbor empirical potentials~\cite{tersoff,edip} have also
been tested, but they have proven to be unsuitable to reproduce the 
complexity of these reconstructions.

Given the small energy difference between the two reconstructions, 
we have checked if at finite temperatures one of the two phases prevailed 
more clearly. We have calculated the Helmholtz free energy $F$ following a
quasi-harmonic approach up to 300~K~\cite{free_energy}. We have found that 
the difference between the two structure remains practically
unchanged all over the temperature range analyzed. This is due to
the fact that the vibrational contribution to $F$ is determined by
the integrated phonon density of states which hardly changes
from one case to the other.

If SiNWs are to be used as nanoswitches or for manipulation purposes, 
it is important that they have a certain stiffness that prevents them 
from collapsing as a result of mechanical tensions. We have studied 
the response to axial stress in $\langle 100 \rangle$ SiNWs within 
non-orthogonal TB, finding a Young's modulus of $\sim$~137~GPa 
($\sim$~195~GPa for bulk Si) and a Poisson ratio of $\sim$~0.35~\cite{young}. 
These values confirm the intuition that their bulk core make SiNWs 
mechanically stable.

\begin{figure}[h]
\begin{center}
\epsfxsize=8cm
\epsffile{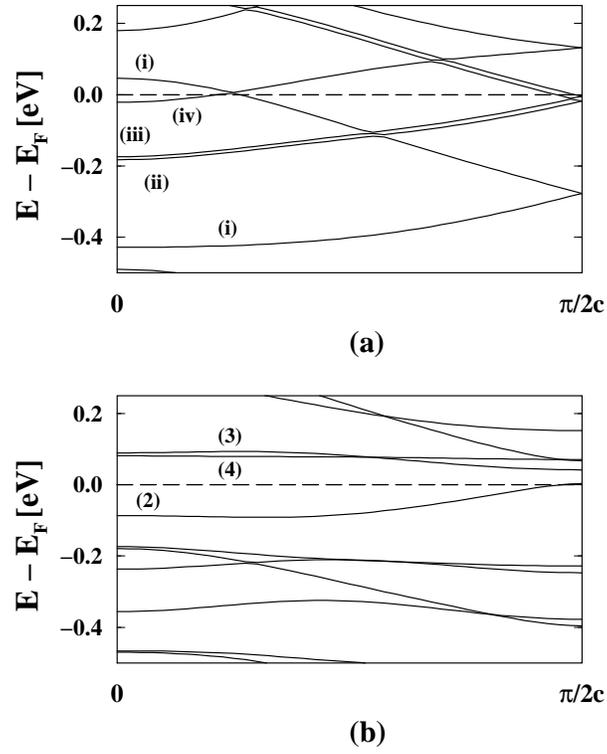}
\end{center}
\caption{Band structure diagrams. (a) The $1c$ reconstruction
         has four metallic states [(i) to (iv)] clearly crossing the
         Fermi level; (b) the $2c$ reconstruction has rather a
         single semi-metallic state.
         For comparison, both diagrams refer to the $2c$ unit cell.}
\label{fig:bands}
\end{figure}

\begin{figure}[h]
\begin{center}
\epsfxsize=8.5cm
\epsffile{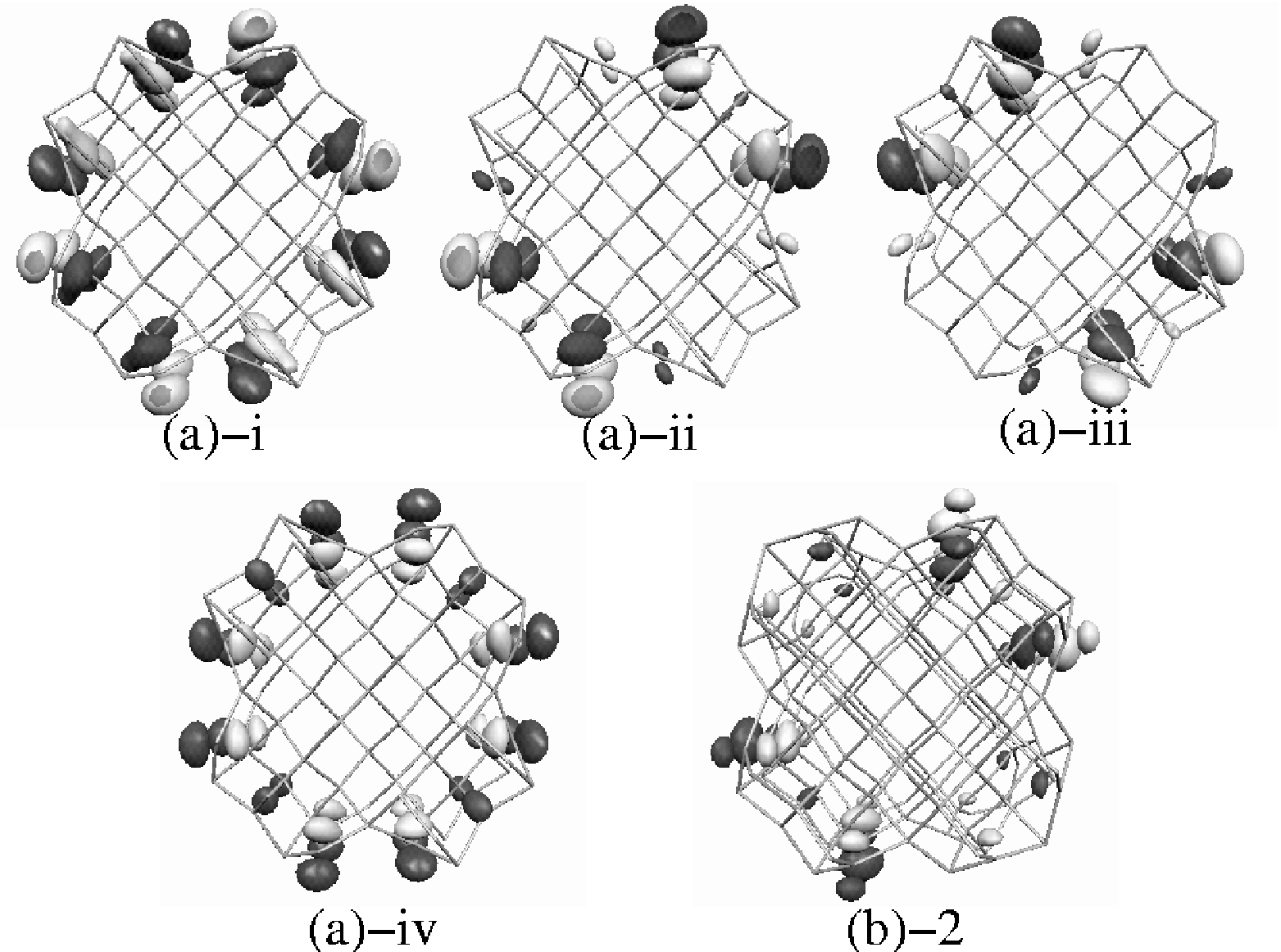}
\end{center}
\caption{Cross-section view of the wave functions at $\Gamma$ point of
         (a) the four metallic states of the $1c$ reconstruction
         [(i) to (iv)]  and
         (b) the semi-metallic state of the $2c$ reconstruction.
         Labeling follows the scheme of Fig.~(\ref{fig:bands}).
        These wave functions give a false impression of C$_{2v}$ symmetry. The
         {\em pseudo} C$_{2v}$ symmetry explains why bands $(ii)$ and $(iii)$ are
        almost degenerate, (surface states $(a)-ii$ and $(a)-iii$ in the figure).}
\label{fig:wf}
\end{figure}

We have calculated the band structure diagrams corresponding to the 
two different reconstructions. Surprisingly, the electronic 
structures turn out to be utterly different, especially if compared to 
their rather similar geometries. The $1c$-reconstruction has four 
clearly metallic states [see Fig.~(\ref{fig:bands}a)], while the $2c$ 
wire shows a semi-metallic behavior with a single Fermi-level crossing 
at the zone boundary [see Fig.~(\ref{fig:bands}b)]. The metallic character 
of the $1c$-wire is robust against failures of DFT. A scissor operator 
correction would rigidly shift some of the electronic bands. In order to 
open a gap, both of the degenerate bands~(ii) and (iii) in Fig.~(\ref{fig:bands}a) 
should become fully occupied or fully empty because both bands have the same
symmetry (see Fig.~(\ref{fig:wf})). Given the band structure about the Fermi 
energy, it is not possible to find shifts of other bands to prevent this shift 
from violating electron counting. Thus the $1c$-wire has metallic character.
However, the limitations of DFT approximations to give the correct band gap 
of bulk silicon only permit to claim that the $2c$-wire is a semi-metal of 
a small gap material.

The two reconstructions evince two competing mechanisms ruling
the surface energetics in connection with the electronic structure
of the SiNWs. There is a trade off between recovering the bulk-like 
tetrahedral coordination on one side and the full delocalization of 
a genuine Bloch state along the wire axis on the other side. 
The first of these two mechanisms favors the flipping of one of the 
dimers which is characteristic of the $2c$-reconstruction. 
The larger coordination of the protruding Si atom in the
flipped dimer leads to angles of $\sim 100^\circ$, while the corresponding 
angles in the $1c$-reconstruction are of $\sim 90^\circ$. This energy 
gain is compensated by the shift of the electronic structure as shown in 
Fig.~(\ref{fig:bands}). This shift is due to the reduction of symmetry 
by the flipped dimer. The flipped dimer interrupts the surface Bloch 
states, hence the overlap between dangling bonds is reduced in the 
$2c$-reconstruction and the energy of the surface states increases. 
Electronic localization induces an increase of energy by breaking the
one-dimensional Bloch state, but on the other hand makes the system
lower its energy locally by increasing the bonding of one of the dimers. 

The bands of the $1c$-reconstruction can be identified with the
$2c$-reconstruction ones, Fig.~(\ref{fig:bands}). The $(ii)$ band of
Fig.~(\ref{fig:bands}a) shows similar symmetry as the $(2)$ of
Fig.~ (\ref{fig:bands}b) (see Fig.~\ref{fig:wf} below). An adiabatic
calculation, slowly flipping the dimer yields this one-to-one correspondence.
In the same manner, bands $(iii)$ and $(iv)$ of Fig.~(\ref{fig:bands}a) 
can be identified with $(4)$ and $(3)$ of Fig.~(\ref{fig:bands}b) 
respectively. The effect on the electronic structure of flipping one 
dimer is to localize, leading to higher-energy and flatter bands, 
Fig.~(\ref{fig:bands}b).

The electronic structure about the Fermi energy is given by the surface 
states originating in the dimer dangling bonds. This is shown in 
Fig.~(\ref{fig:wf}). Iso-surfaces of wave-function amplitude have been 
plotted on the SiNW atomic structure, showing that Bloch states originate 
on the facet dangling bonds. 
Figure~(\ref{fig:wf}) also shows the correspondence of the wave 
functions (a)-ii and (b)-2, as explained above. 
Surface states are thus present in both reconstructions.  

Conductance through thin SiNWs remains a difficult issue~\cite{heath}. 
The way SiNWs bind to metal electrodes is complex and crucial for 
determining the actual conductance of SiNW-based devices. In the absence 
of defects and in the case of excellent electrode contacts, the strongly 
metallic SiNW ($1c$-reconstructed) will present a maximal conductance 
of four quanta, equivalent to a resistance of only $ \sim 3 k \Omega$ per 
SiNW. In the same conditions the semi-metallic SiNW ($2c$-reconstructed) 
exhibits a minimum resistance of $ \sim 13 k \Omega$ per SiNW.
These results show that SiNWs can be good conductors, and their actual 
electrical properties will be strongly dependent on the growth conditions.
Experimental studies~\cite{heath} show that despite diffusion 
from the metallic contacts and/or surface contamination the measured 
conductance of thin SiNWs after annealing is much higher than the one 
expected from doping the SiNWs. This finding is in qualitative agreement 
with the above surface-state driven conduction. 

Any perturbation on the surface states will drastically affect the 
conduction properties of the wires. These wires will be extremely 
sensitive to the chemical environment and therefore are good candidates 
for molecular detection~\cite{zhou,cui,hahm}.

In conclusion, we have performed DFT calculations allowing for different 
reconstructions of $\langle 100 \rangle$ SiNWs with a bulk core. The SiNWs 
are classified according to their \{100\}-facet 
reconstructions. No direct correspondence is found with infinite 
surface reconstructions, hence the two minimum-energy possibilities 
are classified according to the size of their periodicity along the 
wire axis: $1c$ for the one with the shortest period and $2c$ for the 
longest one. The bulk-like core of the wire confers them with a Young 
modulus close to the bulk Si one; on the other hand, the Poisson ratio 
indicates good lateral elastic properties of these wires. Despite the 
closeness in energy of the two reconstructions, the $1c$ one shows a 
strong metallic character while the $2c$ one is semi-metallic. We have 
rationalized these findings in terms of the gain in energy due to the 
formation of surface state versus tetrahedral-like angles in the surface 
dimers.  Surface states are probably ubiquitous in a large family of SiNWs. 
The wires studied here present good conduction properties thanks to the 
reconstruction-induced surface states. 
Hence, the use of thin SiNWs as conductors without doping is possible.

R.R. acknowledge the financial support of the Generalitat de Catalunya 
through a {\sc Nanotec} fellowship. N.L. thanks ACI jeunes chercheurs. 
Computational resources at the Centre Informatique National de l'Enseignement 
Sup\'erieur and the Centre de Calcul Midi-Pyr\'en\'ees are gratefully 
acknowledged. We thank A.~G.~Borisov, E.~Hern\'{a}ndez and P.~Ordej\'{o}n
for their suggestions.

\newpage

\newpage

\newpage

\begin{thebibliography}{99}
\bibitem{cohesion} We have obtained a value of 4.444~eV/atom for bulk Si.
\bibitem{nota} According to Ref.~\onlinecite{ramstad}, a numerical 
               uncertainty of a few meV per dimer is to be expected 
               for the Si(100) surface DFT calculations. Our conclusion 
               is then that given the numerical convergency attained in 
               our calculations, the 3~meV/atom difference is significative. 
               The GGA of Ref.~\protect{\onlinecite{gga}} used in the present 
               calculations gives correct geometries and accurate total 
               energies for surfaces. In particular recent studies of Si(100) 
               show that the typical approximations of DFT give correct 
               results in surface reconstruction and geometries (see 
               Ref.~\onlinecite{healey}). Hence, the reconstructions 
               presented here are probably within a few percent of the actual 
               bond distances and angles. The metallicity of different 
               Si surface reconstructions is reproduced by DFT calculations.
               Indeed, the  metallic character of a given surface reconstruction 
               has been used to predict and analyze the actual surface 
               reconstructions of semiconductors, see for example 
               Ref.~\onlinecite{chadi}.
\bibitem{young} The Young's modulus is the ratio between stress and the axial 
                strain $\epsilon$ and is conventionally defined as 
                $Y = \frac{1}{V_0} \frac{\partial^2E}{\partial\epsilon^2} 
                |_{\epsilon=0}$, where $V_0$ and $E$ are the equilibrium 
                volume and the total energy respectively; the Poisson ratio 
                is defined as 
                $\sigma = - \frac{1}{\epsilon} \frac{R-R_0}{R_0}$,
                where $R_0$ is the equilibrium radius of the SiNW 
                and $R$ is the radius at strain $\epsilon$.
\end{thebibliography}
\end{document}